\documentstyle[12pt]{article}
\newcommand{\z}{&\hspace*{-8pt}}
\newcommand{\as}{\alpha_{\rm s}}

\begin{document}

\vspace*{1cm}
\begin{center}
{\Large \bf Quark Mass Corrections to the Bjorken and\\
Gross--Llewellyn-Smith Sum Rules}\\[20mm]
{\bf \large O.V. Teryaev and O.L. Veretin}\\
{\it Bogoliubov Laboratory of Theoretical Physics,\\
JINR, Dubna 141980, Russia}
\end{center}
\vskip 20mm

\begin{abstract}
  Quark mass corrections to the spin partonic structure function
$g_1(x,Q^2)$ and function $F_3(x,Q^2)$ are obtained at the order $O(\as)$
along with the coefficient
functions $C^{(A)}$ and $C^{(V)}$ related to the Bjorken and
Gross--Llewellyn-Smith sum rules.
In the mass\-less li\-mit the dif\-ference
be\-t\-we\-en $F_3$ and $g_1$ is en\-co\-unte\-red
to be $(\as/\pi)C_F(1-x)$. The results for the functions
$C^{(A)}$ and $C^{(V)}$ at $m=0$  agree with the previous MS-scheme
calculations $C^{(A)}=C^{(V)}=1-(3\as/4\pi)C_F$.
\end{abstract}

\newpage

  Recently many theoretical efforts were spent to study
radiative corrections to the polarized Bjorken
sum rule (BSR) \cite{Bjorken} in deep inelastic $eN$-scattering
\begin{eqnarray}
  \int\limits_0^1 \left( g_1^{ep}(x,Q^2)-g_1^{en}(x,Q^2) \right)
     {\rm d}x = \frac13 \left|\frac{g_A}{g_V}\right| C^{(A)},
		\label{BSR}
\end{eqnarray}
where $g_A$ and $g_V$ are the constants in nuetron weak decay and
$C^{(A)}$ is the coefficient function of the axial current
in the operator product expansion of two vector currents.
Corrections to $C^{(A)}$ are known to coincide with those to
another coefficient function $C^{(V)}$ which is relevant for the
Gross--Llewellyn-Smith (GLS) sum rule \cite{Gross} in neutrino-nucleon
scattering
\begin{eqnarray}
  \int\limits_0^1 \left( F_3^{\bar\nu p}(x,Q^2) + F_3^{\nu p}(x,Q^2) \right)
                     {\rm d}x = 3 C^{(V)}.
		\label{GLS}
\end{eqnarray}

   The leading corrections to BSR and GLS were first computed in
Refs. \cite{Kodaira,Bardeen} and appeared to be $C^{(A)}=C^{(V)}
=1-C_F(3\as/4\pi)$. Next to leading order results for both
sum rules can be found in Ref. \cite{Gorishny}. Discrepancies
between $C^{(A)}$ and $C^{(V)}$ arise only at order $O(\as^3)$
\cite{Larin} where 'light-by-light' disgrams appear.
These calculations were performed in MS scheme with quark mass $m=0$.
From the other hand recent time mass dependent RG equations
\cite{Shirkov} and power corrections \cite{Karliner}
attract a great attention.
Thus it is of interest to get mass dependence of coefficient functions.

  In the recent work \cite{Mertig} there were computed $O(\as)$
corrections to $C^{(A)}$ in on-shell scheme. It was noticed that
there are contributions to the partonic structure function $g_1$
that survive when $m\to0$. That is the diagram of the box type which is
responsible for that. After momentum integration it develops an additional
compensating factor $1/m^2$ which cancels with the mass in
numerator resulting in finite terms.
Thus the result for $g_1$ is different in dependence on
whether mass $m=0$ from the very beginning or it is kept till the end of
calculations.

  Below the partonic structure functions both $g_1$ and $F_3$
at order $O(\as)$ with $m\not=0$ will be presented.
Along this paper we use the on-shell renormalization scheme.

  First we note that the contribution of virtual gluon
can be expressed in terms of renormalized
elastic formfactors of currents. Corrections to the vector (axial)
current are usually written as ($q=p'-p$)
\begin{eqnarray}
  V_\mu \z=\z \bar u(p')\left\{
    \gamma_\mu \Bigl( 1+\frac{\as}{\pi}C_F{\cal F}_1^V(q) \Bigr)
       + \frac{1}{4m} [\hat q,\gamma_\mu]_{-}
         \frac{\as}{\pi}C_F{\cal F}_2^V(q) \right\} u(p),
	      \label{V vertex}  \\
  A_\mu \z=\z \bar u(p')\left\{
  \gamma_\mu\gamma_5 \Bigl( 1+\frac{\as}{\pi}C_F{\cal F}_1^A(q) \Bigr)
       + \frac{1}{4m} (p+p')_\mu \gamma_5
         \frac{\as}{\pi}C_F{\cal F}_2^A(q) \right\} u(p).
	  \label{A vertex}
\end{eqnarray}
   These formfactors were computed earlier in QED and electroweak theory
(see e.g. \cite{formfactors} and references therein).
We take them in the following form
\begin{eqnarray}
  {\cal F}_1^V \z=\z -\left( 1+\frac{1+\theta^2}{1-\theta^2}\log\theta
        \right) \log\frac{\mu}{m} - 1 -
	\frac{3\theta^2+2\theta+3}{4(1-\theta^2)} \log\theta  \nonumber\\
   \z+\z \frac{1+\theta^2}{1-\theta^2} \left( -\frac14\log^2\theta
      +\frac12\zeta(2)+{\rm Li}_2(-\theta)
       +\log\theta\log(1+\theta)\right),
                      \label{F1V}  \\
  {\cal F}_2^V \z=\z - \frac{\theta}{1-\theta^2}\log\theta,
		      \label{F2V}  \\
       \label{formfactors}
   {\cal F}_1^A \z=\z {\cal F}_1^V +
        \frac{\theta}{1-\theta^2}\log\theta,
		      \label{F1A}
\end{eqnarray}
with
\begin{eqnarray}
  \theta \z=\z \frac{\sqrt{1+4r}-1}{\sqrt{1+4r}+1},\qquad r=\frac{m^2}{Q^2}.
	        \label{definitions}
\end{eqnarray}
  In the above formulae $\mu$ is a small 'gluon mass' ($\mu\ll t,m^2$)
being infrared regulator and $m$ is a quark mass.
Formfactor ${\cal F}_2^A$ is irrelevant and hence is omitted.
The vector current is normalized such that ${\cal F}_1^V(q=0)=1$. If
such the condition is imposed then for the axial current one gets
${\cal F}_1^A(0)=1-(\as/2\pi)$. Formfactor ${\cal F}_2^V(0)=1/2$
is the well known anomalous magnetic moment. Let us emphasize that it is
impossible to normalize both currents to unity in the presence of a
nonvanishing mass. We shall return to this question later on.

  Using definitions (\ref{V vertex}),(\ref{A vertex}) the virtual
contributions can be cast into the form
\begin{eqnarray}
  g_1^{\rm virtual} \z=\z \frac12\frac{\as}{4\pi}C_F
     \delta(1-x) \left\{ 8{\cal F}_1^V + 4{\cal F}_2^V \right\},
	     \label{g1 virtual}  \\
  F_3^{\rm virtual} \z=\z \frac12\frac{\as}{4\pi}C_F
     \delta(1-x) \left\{ 4{\cal F}_1^V + 4{\cal F}_1^A \right\}.
	     \label{F3 virtual}
\end{eqnarray}

  Next we turn to the real gluon contributions.
Again one faces with IR divergencies
which are due to soft or collinear gluon emission.
As above they are regularized
by letting a gluon have a small nonzero mass $\mu$. Calculating
strightforwardly the diagrams one obtains functions which have
a nontrivial dependence on $\mu$. It causes singularities like $1/(1-x)$
at the end point of $x$-integration (in presence of a small mass $\mu$
$x$ is varying in the interval $0<x<1-2r\mu/m$).
Of course after integration over
variable $x$ the $\mu$ dependence is cast into that as in formulae
(\ref{g1 virtual}),(\ref{F3 virtual}) with the opposite sign so that
$\mu$ cancels in the whole moments. To cancel out $\mu$ before
integration we rewrite result using the well known 'plus distribution'
\cite{plus distribution}, i.e. every function $f(x)$ being singular
at $x=1$ is replaced by
\begin{equation}
  f(x) = f_+(x)
    + \delta(1-x) \!\!\! \int\limits_0^{1-2r\mu/m}\!\!\! f(z) \,{\rm d}z,
    \label{singular}
\end{equation}
where the limit $\mu\to0$ is implied. After some transformations
the structure functions can be presented as
\begin{eqnarray}
   w^{\rm real}(x) \z=\z
    \frac12\frac{\as}{4\pi} C_F
       \left\{ \delta(1-x){\cal R}
      + 4(1+2r)\left(\frac{L}{1-x}\right)_{\!\!\!+} - \frac{8}{(1-x)_+}
      + c_1 + c_2 L \right\},
	   \label{w real}   \\
  L \z=\z \frac{1}{\sqrt{1+4rx^2}}
    \log\frac{1+2rx+\sqrt{1+4rx^2}}{1+2rx-\sqrt{1+4rx^2}}.
\end{eqnarray}
Here $w(x)$ stands for either $g_1$ or $F_3$. All infrared divergencies
now are absorbed in coefficients ${\cal R}$. By explicit calculation we
have found for ${\cal R},c_1$ and $c_2$
\begin{eqnarray}
  {\cal R}^{g_1} \z=\z - 8{\cal F}_1^V - 4{\cal F}_2^V
    - 4 + 8\log r - 2\frac{5\theta^2+4\theta+5}{1-\theta^2}\log\theta
                   \label{R g1}  \\
  {\cal R}^{F_3} \z=\z - 4{\cal F}_1^V - 4{\cal F}_1^A
    - 4 + 8\log r - 10\frac{1+\theta^2}{1-\theta^2}\log\theta
                   \label{R F3}  \\
  c_1^{g_1} \z=\z \frac{1}{(1+4rx^2)^2(1-x+rx)^2}
   \Bigl[  (1-x)(3+6x-8x^2)   \nonumber\\
    \z+\z 4rx(1-x)(2+15x-15x^2+4x^3)  \nonumber\\
    \z+\z 4r^2x^2(1+24x-29x^2+6x^3) + 8r^3x^4(5-x)  \Bigr]
                \label{c1 g1}    \\
  c_1^{F_3} \z=\z \frac{1}{(1+4rx^2)(1-x+rx)^2}
   \Bigl[ (1-x)(7-9x-2x-4x^2)  \nonumber\\
    \z+\z 2rx(7-3x-4x^3) + 8r^2x^2 \Bigr]
	        \label{c1 F3}   \\
  c_2^{g_1} \z=\z \frac{-2}{(1+4rx^2)^2}
    \Bigl[ 1+x + 2r(2+x+9x^2-4x^3) \nonumber\\
    \z+\z 4r^2x^2(8+5x-x^2) + 64r^3x^4 \Bigr]
                \label{c2 g1}  \\
  c_2^{F_3} \z=\z \frac{-2}{1+4rx^2}
    \Bigl( 1+x + 4r(1+x) + 16r^2x^2 \Bigr)
		\label{c2 F3}
\end{eqnarray}
  From the formulae (\ref{g1 virtual}),(\ref{F3 virtual}) and
(\ref{w real})--(\ref{R F3}) one can see that $\log\mu$'s cancel as 
well as ${\cal F}_{1,2}^{A,V}$ and we are left with IR safe expressions.

  If $Q^2 \gg m^2$ then the formulae for $g_1,F_3$ greatly simplify.
Adding the Born contribution and keeping only leading terms in $m^2/Q^2$
we arrive at
\begin{eqnarray}
  g_1(x) \z=\z \frac12\delta(1-x)+
   \frac12\frac{\as}{4\pi}C_F
   \Bigl\{ 2 \Bigr[ \frac32\delta(1-x)
   +\frac{1+x^2}{(1-x)_+}\Bigr] \log\frac{Q^2}{m^2} - 5\delta(1-x)
                      \nonumber\\
   \z-\z \frac{7}{(1-x)_+}
   - 4\left(\frac{\log\bigl(x(1-x)\bigr)}{1-x}\right)_{\!\! +}
   + 2 + 8x + 2(1+x)\log\bigl(x(1-x)\bigr)  \Bigr\}
                  \label{g1 massless}    \\
  F_3(x) \z=\z \frac12\delta(1-x)+
   \frac12\frac{\as}{4\pi}C_F
   \Bigl\{ 2 \Bigr[ \frac32\delta(1-x)
   +\frac{1+x^2}{(1-x)_+}\Bigr] \log\frac{Q^2}{m^2} - 5\delta(1-x)
                      \nonumber\\
   \z-\z \frac{7}{(1-x)_+}
   - 4\left(\frac{\log\bigl(x(1-x)\bigr)}{1-x}\right)_{\!\! +}
   + 6 + 4x + 2(1+x)\log\bigl(x(1-x)\bigr)  \Bigr\}
                  \label{F3 massless}    \\
\end{eqnarray}

   As it was mentioned in the beginning of the paper corrections
to the structure function $F_3$ evaluated in massless theory
coincide with those to $g_1$.
Here we see from (\ref{g1 massless}),(\ref{F3 massless}) that this is
not true in massless limit of the massive formulae. The difference is
totally due to coefficients $c_1$'s. It is worth noting that the result
obtained should not depend on an infrared regularization procedure
provided quark mass $m$ tends to zero after IR singularities
are canceled out. Eqns. (\ref{g1 massless}),(\ref{F3 massless}) yield
that $g_1$ develops an extraterm $-C_F(\as/2\pi)(1-x)$. In fact
this term defines a fermion helicity-flip probability $P_{+-}(x)$ and was
studied in Ref. \cite{Falk}.

  Using formulae (\ref{R g1})--(\ref{c2 F3}) we obtain first moments
of $g_1$ and $F_3$ with no approximation made
\begin{eqnarray}
  2\int\limits_0^1 w(x) \,{\rm d}x = 1+\frac{\as}{4\pi}C_F\delta,
	  \qquad w(x) = g_1(x),F_3(x).
	       \label{w moment}
\end{eqnarray}
  Where the factors $\delta$'s read
\begin{eqnarray}
  \delta^{g_1} \z=\z
  - 4 + \frac{1-16r}{2}I(r) - \frac{2-17r+16r^2}{2r(1-r)}\log r
      +\frac{1-5r-28r^2}{r\sqrt{1+4r}}\log\theta,
                  \label{g1 delta}  \\
  \delta^{F_3} \z=\z
  -\frac{5-4r}{1-r} - 8r I(r) + \frac{1+6r-16r^2+8r^3}{r(1-r)}\log r
     - \frac{1+10r+24r^2}{r\sqrt{1+4r}}\log\theta.
                  \label{F3 delta}
\end{eqnarray}
  Function $I(r)$ can be expressed through Euler dilogarithm function
${\rm Li}_2$ and has the following representation
\begin{eqnarray}
  I(r) \z=\z \int\limits_0^1 \frac{{\rm d}x}{\sqrt{1+4rx^2}}
     \log\frac{1+2rx+\sqrt{1+4rx^2}}{1+2rx-\sqrt{1+4rx^2}} \nonumber \\
   \z=\z \frac{1}{2\sqrt{r}} \bigl\{
       {\rm Li}_2(-t) - {\rm Li}_2(t) + {\rm Li}_2(-at)
      + {\rm Li}_2(\frac{a}{t}) - \frac12 {\rm Li}_2(a^2)
      + \frac{\pi^2}{4} + \frac12\log^2 t \bigl\},
		  \label{I}  \\
   \z\z a = (1-\sqrt{r})/(1+\sqrt{r}), \qquad
	t = \sqrt{1+4r}-2\sqrt{r}.
\end{eqnarray}
  In high $Q^2$ region it turns to be
$I(r) = 2 - \log r + r \bigr( 2/9+(5/3)\log r \bigl)+O(r^2\log r)$.

  Eqns. (\ref{w moment})--(\ref{F3 delta}) together with (\ref{I})
give the values of the first moments of the partonic structure
functions in the whole region of $Q^2>0$.
In deep inelastic case ($m^2/Q^2 \to 0$) we obtain
$g^{(1)}_1=1-C_F(5\as/4\pi)$ while within the massless approach
it is $1-C_F(3\as/4\pi)$. This result was found earlier
in Ref. \cite{Mertig}.

  Let us consider now the coefficient functions. The OPE says that
\begin{eqnarray}
  2 g^{(1)}_1(Q^2) \z=\z C^{(A)}(Q^2) A^{(1)}_5, \\
  2 F^{(1)}_3(Q^2) \z=\z C^{(V)}(Q^2) A^{(1)},
\end{eqnarray}
where $A$'s are defined from operator matrix elements
\begin{eqnarray}
  \langle p,s|\bar \psi(q)\gamma_5\gamma_\mu \psi(q)|p',s \rangle  \z=\z
	    \bar u(p)\gamma_\mu\gamma_5 u(p') A^{(1)}_5(t),
                       \label{A15}\\
  \langle p|\bar \psi(q)\gamma_\mu \psi(q)|p' \rangle  \z=\z
	    \bar u(p)\gamma_\mu u(p') A^{(1)}(t), \label{A1}\\
			  t \z=\z (p'-p)^2,
\end{eqnarray}
with quark field $\psi$. Matrix elements (\ref{A15}),(\ref{A1})
must be taken at zero momentum transfer $t=0$ in the on-shell scheme.
For the vector current
the identity $A^{(1)}\equiv 1+(\as/\pi){\cal F}_1^V(t=0)$
follows and it is equal to unity because of the current normalization
while for the axial we have $A^{(1)}_5(t)\equiv
1+(\as/\pi)\bigl( {\cal F}_1^A(t) - {\cal F}_1^V(t) \bigr)$. The
structure in the parentheses looks like
\begin{eqnarray}
 {\cal F}_1^A(t) - {\cal F}_1^V(t)
        = \frac{\theta}{1-\theta^2}\log\theta,
	  \label{F1A-F1V}
\end{eqnarray}
with $\theta$ defined as in (\ref{definitions}) and $r=m^2/t$.
For large $t$ (\ref{F1A-F1V}) vanishes as it should be due to
chiral invariance. This situation is realized when $m$ is identically equal
to zero, when the limit $t \to 0$ corresponding to the forward matrix
element could be easily taken. However, if one take this limit {\it before}
setting $m=0$ the mass terms come
into the game and (\ref{F1A-F1V}) becomes $-1/2$. One can check
this using (\ref{F1V})--(\ref{definitions}).  As a result $C^{(A)}$
differs form the $g_1^{(1)}$ at $m=0$ by a finite term
\begin{eqnarray}
  C^{(A)} = \left( 1 - \frac{5\as}{4\pi}C_F \right)
            \left( 1 - \frac{\as}{2\pi}C_F  \right)^{-1}
        = 1 - \frac{3\as}{4\pi}C_F + O(\as^2).
\end{eqnarray}

  Let us summarize now the results. Coefficient functions $C$'s with a
nonvanishing fermion mass look like
\begin{eqnarray}
   C^{(A)} \z=\z 1 + \frac{\as}{4\pi} C_F (\delta^{g_1}+2),\label{C(A)}\\
   C^{(V)} \z=\z 1 + \frac{\as}{4\pi} C_F \delta^{F_3},    \label{C(V)}
\end{eqnarray}
where functions $\delta$ are given by (\ref{g1 delta}),(\ref{F3 delta}).
In Fig.1 there are the plots of the corrections to
coefficient functions $C^{(A)},C^{(V)}$
versus $m^2/Q^2$. In the deep inelastic limit both corrections
coincide with each other in agreement with the values quoted in literature.

  Up to the terms $O(m^2/Q^2)^2$ the Eqns. (\ref{C(A)}),(\ref{C(V)}) read
\begin{eqnarray}
   C^{(A)}(Q^2) \z=\z 1 - \frac{\as}{4\pi} C_F
     \left[ 3+\frac{m^2}{Q^2}
      \Bigl( \frac{11}{3}\log\frac{m^2}{Q^2} - \frac{10}{9} \Bigr) \right],\\
   C^{(V)}(Q^2) \z=\z 1 - \frac{\as}{4\pi} C_F
     \left[ 3+\frac{m^2}{Q^2}
      \Bigl( 3\log\frac{m^2}{Q^2} + 4 \Bigr) \right].
\end{eqnarray}
  The discrepancy between $C^{(A)}$ and $C^{(V)}$ manifests a violation
of the Crewther relation \cite{Crewther} by mass corrections.

  The magnitude of the strange quark mass correction is about $0.12$
of the massless one-loop result at $Q^2=2\,{\rm GeV}^2$.
For the light quarks, the mass
contribution is negligible if one uses the current quark mass of order of
few MeV. If, however, one takes into account that such a scale should
be non-observable and substitute instead the scale of order
of a pion mass \cite{Gorsky}, the result is still about $0.1$.

  The calculated corrections are in fact the first example of the
NLO mass dependence in QCD. Generally speaking this would require
to calculate the 2-loop mass-dependent anomalous dimension and one-loop
coefficient function. However in the case at hand the
anomalous dimension is zero and the calculated contribution provides
the final result.

  After this work was completed, the paper \cite{Neerven2} 
appeared, where the coefficient functions for heavy quarks are
investigated in the limit $Q^2 \geq m^2$.

  We are indebted to A.L. Kataev, S.V. Mikhailov and I.V. Musatov for useful
discussions. O.T. is grateful to J. Collins 
for elucidating correspondence and to W. van Neerven for valuable comments.
  
  This work was supported by RFFR grant N 93-02-3811.


\input epsf.sty     
\epsfverbosetrue

\begin{minipage}[t]{140mm}

\begin{center}
\hspace*{-1.5cm}
\mbox{
   \epsfxsize=13.0cm                    
   \epsfbox[0 0 700 700]{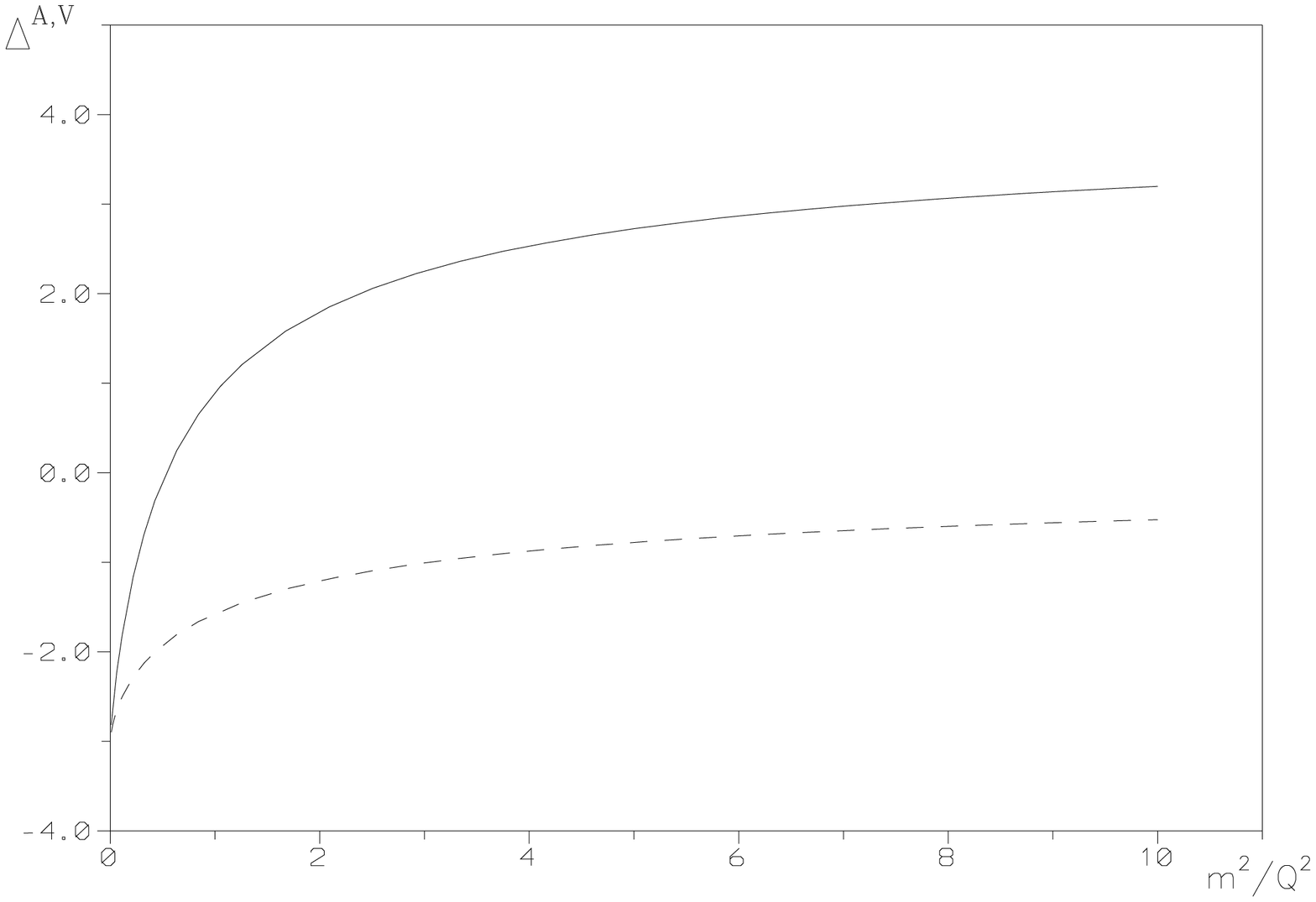}       
     }
\end{center}

\vspace{-0.5cm}
{\bf Fig.~1.}
Corrections to the Bjorken (solid line) and
Gross--Llewellyn-Smith (dashed line)
sum rules versus $m^2/Q^2$. Coefficient functions are written
in the form $C^{(A),(V)}=1+(\as/4\pi)C_F \Delta^{A,V}$.

\end{minipage}

\end{document}